 \documentclass{emulateapj}

\usepackage{apjfonts}

\newcommand{\object}[1]{#1}

\slugcomment{to appear in the Astrophysical Journal, Letters}

\shorttitle{Hydrogen Burning Turnoff of RS Oph 2006}
\shortauthors{Hachisu et al.}

\begin{document}


\title{The Hydrogen Burning Turn-off of RS Ophiuchi 2006}


\author{Izumi Hachisu\altaffilmark{1},
Mariko Kato\altaffilmark{2},
Seiichiro Kiyota\altaffilmark{3},
Katsuaki Kubotera\altaffilmark{4},
Hiroyuki Maehara\altaffilmark{5},
Kazuhiro Nakajima\altaffilmark{6},
Yuko Ishii\altaffilmark{7},
Mari Kamada\altaffilmark{7},
Sahori Mizoguchi\altaffilmark{7},
Shinji Nishiyama\altaffilmark{7},
Naoko Sumitomo\altaffilmark{7},
Ken'ichi Tanaka\altaffilmark{7},
Masayuki Yamanaka\altaffilmark{7},
and Kozo Sadakane\altaffilmark{7}}

\altaffiltext{1}{Department of Earth Science and Astronomy,
College of Arts and Sciences,
University of Tokyo, Komaba 3-8-1, Meguro-ku, Tokyo 153-8902,
Japan: hachisu@chianti.c.u-tokyo.ac.jp}
\altaffiltext{2}{Department of Astronomy, Keio University, 
Hiyoshi 4-1-1, Kouhoku-ku, Yokohama 223-8521, Japan:
mariko@educ.cc.keio.ac.jp}
\altaffiltext{3}{Azuma 1-401-810, Tsukuba 305-0031, Japan:
skiyota@nias.affrc.go.jp}
\altaffiltext{4}{Numashiro 702-3, Odawara 256-0801, Japan:
k\_kubotera@nifty.com}
\altaffiltext{5}{Department of Astronomy, School of Science, 
University of Tokyo, Hongo 7-3-1, Bunkyo-ku, Tokyo 113-0033, Japan:
maehara@provence.c.u-tokyo.ac.jp}
\altaffiltext{6}{Teratani 124, Isato, Kumano, Mie 519-4673, Japan:
k.nakajima@ztv.ne.jp}
\altaffiltext{7}{Astronomical Institute, Osaka Kyoiku University,
Kashiwara-shi, Osaka 582-8582, Japan: sadakane@cc.osaka-kyoiku.ac.jp}




\begin{abstract}
We report a coordinated multi-band photometry of the \object{RS Oph}
2006 outburst and highlight the emission line free $y$-band photometry that
shows a mid-plateau phase at $y \sim 10.2$ mag from day 40 to day 75
after the discovery followed by a sharp drop of the final decline.  
Such mid-plateau phases are observed in other two recurrent novae, 
\object{U Sco} and \object{CI Aql}, and are interpreted as a bright
disk irradiated by the white dwarf.  We have calculated
theoretical light curves based on the optically thick wind theory
and have reproduced the observed light curves including the mid-plateau
phase and the final sharp decline.  This final decline is identified
with the end of steady hydrogen shell-burning, which turned out
the day $\sim 80$.  This turnoff date is consistent with the end of a
supersoft X-ray phase observed with {\it Swift}.
Our model suggests a white dwarf mass of $1.35 \pm 0.01 ~M_\sun$,
which indicates that RS Oph is a progenitor of Type Ia supernovae.
We strongly recommend the $y$-filter observation of novae to detect
both the presence of a disk and the hydrogen burning turn-off.
\end{abstract}





\section{Introduction and Summary}
     \object{RS Oph} is one of the well-observed recurrent novae
and is suggested to be a progenitor system of Type Ia supernovae
\citep[e.g.,][]{hac01kb}.  It has undergone its sixth recorded
outburst on 2006 February 12 UT \citep{nar06}. 
The five previous recorded outbursts occurred
in 1898, 1933, 1958, 1967, and 1985.  These short $10-20$ yr recurrence
periods indicate that the white dwarf (WD) mass is very close to the
Chandrasekhar mass and that the mass accretion rate is as large as 
$\dot M_{\rm acc} \sim 1 \times 10^{-7} M_\sun$~yr$^{-1}$
\citep[see, e.g., Fig.2 of][]{hac01kb}.
If the WD mass increases after each outburst, \object{RS Oph}
will eventually explode as a Type Ia supernova 
\citep[e.g.,][]{nom82, hkn96, hkn99, hac01kb}.
Therefore, it is important to estimate the WD mass
and the accreted mass left on the WD after the outburst.

It is well known that the X-ray turnoff time is a good indicator of the 
WD mass \citep[e.g.,][]{hac05k,hac06ka}.
When the hydrogen shell-burning atop the WD extinguishes, 
a supersoft X-ray phase ends \citep[e.g.,][]{kra96}.  In 
a visual light curve, however, this turnoff is not clear because 
many strong emission lines contribute to it.  To avoid such 
contamination to the continuum flux, we have observed RS Oph with
the Str\"omgren $y$-band filter.  The $y$-filter is an intermediate
bandpass filter designed to cut the strong emission lines in the wide
$V$ bandpass filter, so that its light curve represents the continuum
flux of novae \citep[e.g.,][]{hac06kb}.  We have further modeled
the light curve of RS Oph and have determined the WD mass by fitting 
our modeled light curve with the observation.

     Section \ref{observation_rs_oph} presents our multi-band 
photometry of the \object{RS Oph} 2006 outburst.  The light curve 
fitting of the observation with our numerical model
are presented in \S \ref{light_curve_model}.
Discussion follows in \S \ref{discussion}.


\begin{deluxetable}{lllll}
\tabletypesize{\scriptsize}
\tablecaption{List of Observers \label{observers}}
\tablewidth{0pt}
\tablehead{
\colhead{name of} & \colhead{location} & \colhead{telescope} & 
\colhead{observed} & \colhead{No. of obs.} \cr
\colhead{observer} & \colhead{in Japan} & \colhead{aperture} & 
\colhead{bands} & \colhead{nights ($y$)} 
}
\startdata
Kiyota & Tsukuba & 25cm & $B,V,y,R_c,I_c$ & 24 \cr
Kubotera & Odawara & 16cm & $B,V,y,R_c$ & ~~8 \cr
Maehara & Kawaguchi & 20/25cm & $B,V,y,R_c,I_c$ & 19 \cr
Nakajima & Kumano & 25cm & $B,V,y,R_c,I_c$ & 55 \cr
OKU\tablenotemark{a} & Kashiwara & 51cm & $V,y$ & 25
\enddata
\tablenotetext{a}{Osaka Kyoiku University team}
\end{deluxetable}

\section{Observation} \label{observation_rs_oph}

Optical observations were started just after the discovery of the 2006
outburst \citep{nar06}.  Each observer and their observational
details are listed in Table \ref{observers}.  We have put a special
emphasis on  the Str\"omgren y-filter to avoid contamination
by the strong emission lines.  These y-filters were made
by Custom Scientific Inc.\footnote{http://www.customscientific.com/}
and distributed to each observer by one of the authors (M. Kato).
Kiyota, Kubotera, Maehara, and Nakajima (VSOLJ members) started observation
on February 13 and obtained 65 nights data for $y$-magnitudes (from
February 17 to July 27).
Osaka Kyoiku University (OKU) team obtained $V$ and $y$ magnitudes
of 25 nights starting from February 17 (until July 14).   
The magnitudes of this object were measured by using the local
standard star, TYC2 5094.92.1 (Kiyota) or TYC2 5094.283.1
(the other observers).
We adapted the brightness and color of ($y= V= 9.57$, $B-V= 0.56$) for
TYC2 5094.92.1 and ($y= V= 9.35$, $B-V=1.23$) for TYC2 5094.283.1 from
Tycho2 catalog.


\begin{deluxetable}{lllll}
\tabletypesize{\scriptsize}
\tablecaption{Parameters of RS Oph\tablenotemark{a}
 \label{parameter_of_rsoph}}
\tablewidth{0pt}
\tablehead{
\colhead{parameter} & \colhead{symbol} & \colhead{25\%} &
\colhead{50\%} & \colhead{100\%} \cr
\colhead{} & \colhead{} & \colhead{efficiency} & 
\colhead{efficiency} & \colhead{efficiency} 
}
\startdata
WD mass & $M_{\rm WD}$ & $1.35 M_\sun$ & $1.35 M_\sun$ & $1.35 M_\sun$ \cr
irradiation & $\eta_{\rm eff}$  & 0.25 & 0.50 & 1.0 \cr
disk size & $\alpha$ & 0.24 ($33~R_\sun$)  & 0.34 ($47~R_\sun$) & 
0.48 ($66~R_\sun$) \cr
size of RG & $\gamma$ & 0.24 ($25~R_\sun$) & 0.34 ($35~R_\sun$) &
0.47 ($48~R_\sun$)  \cr
distance & $d$ & 0.9 kpc & 1.3 kpc & 1.7 kpc
\enddata
\tablenotetext{a}{$M_{\rm RG}= 0.7 M_\sun$, the inclination angle of
the binary $i=33\arcdeg$, the separation
$a= 316.5~R_\sun$, $R_1^*= 138.3~R_\sun$, $R_2^*= 102.5~R_\sun$, 
$\beta= 0.05$, $\nu = 2.0$, $T_{\rm RG}= 3550$~K, the hydrogen burning
turnoff date of day 83, 
and $E(B-V)= 0.73$ are common among all models.}
\end{deluxetable}

\begin{figure}
\epsscale{1.20}
\plotone{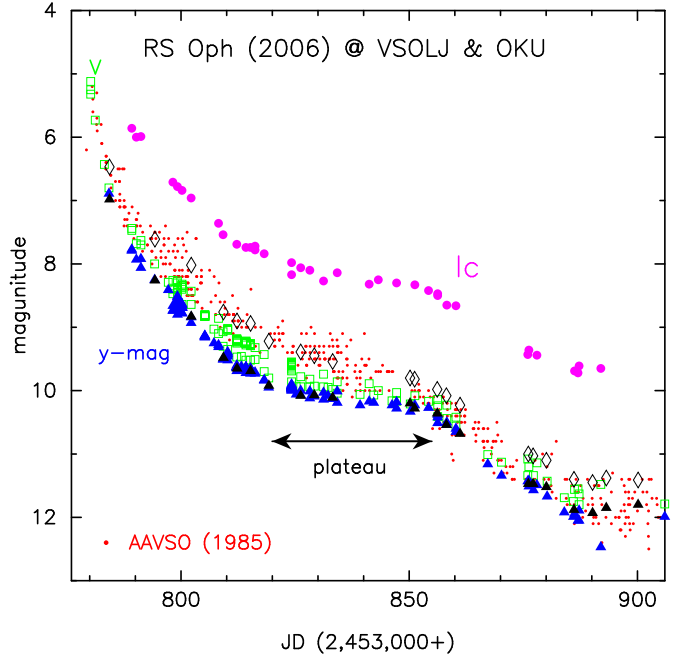}
\caption{
Three ($y$, $V$, and $I_c$) bands light curves for the 2006 outburst
of RS Oph.  The AAVSO visual magnitudes of the previous 1985 outburst
are added for comparison.  We find a plateau phase from day 40
to day 75 and a sharp final decline from day 75.   
{\it Filled circles}: $I_c$-magnitudes.
{\it Filled triangles}: $y$-magnitudes.
{\it Open diamond}: $V$-magnitudes observed by OKU.
{\it Open squares}: $V$-magnitudes observed by Kiyota, Kubotera,
Maehara, and Nakajima (VSOLJ members).
{\it Small dots}: visual magnitudes of the 1985 outburst 
taken from AAVSO.
\label{y_mag_linear}
}
\end{figure}

The $y$-magnitudes are plotted in Figure \ref{y_mag_linear}
together with $I_c$- and $V$-magnitudes.  We have also added  
visual magnitudes of the 1985 outburst from the American Association
of Variable Star Observers (AAVSO) for comparison.  Our $y$-magnitudes
show very small scatter and follows the bottom of the 1985 visual
magnitude. The essential feature of the light curve is very similar
to the previous outbursts.

The $y$ light curve, however, clearly shows a plateau phase
from day 40 to day 75 and the sharp final decline starting from day 75.
Such mid-plateau phases are also observed
in two other recurrent novae, \object{U Sco} \citep[e.g.,][]{hkkm00}
and \object{CI Aql} \citep[e.g.,][]{hac01ka, hac03kb, hac03ka}.
These authors interpreted the mid-plateau phase as a bright disk irradiated
by the hydrogen-burning WD and a sharp start of the final decline
as the epoch when the hydrogen shell-burning ends.


\section{Light Curve Model} \label{light_curve_model}


Our binary model, essentially the same as that in \citet{hac01kb}
except for the free-free emission model (see below) in the early phase
of the outburst, consists of a red giant (RG) star, which is not
filling its Roche lobe, a white dwarf (WD), and a disk around the WD.
A circular orbit with the ephemeris given by \citet{fek00} is assumed.


\begin{figure}
\epsscale{1.20}
\plotone{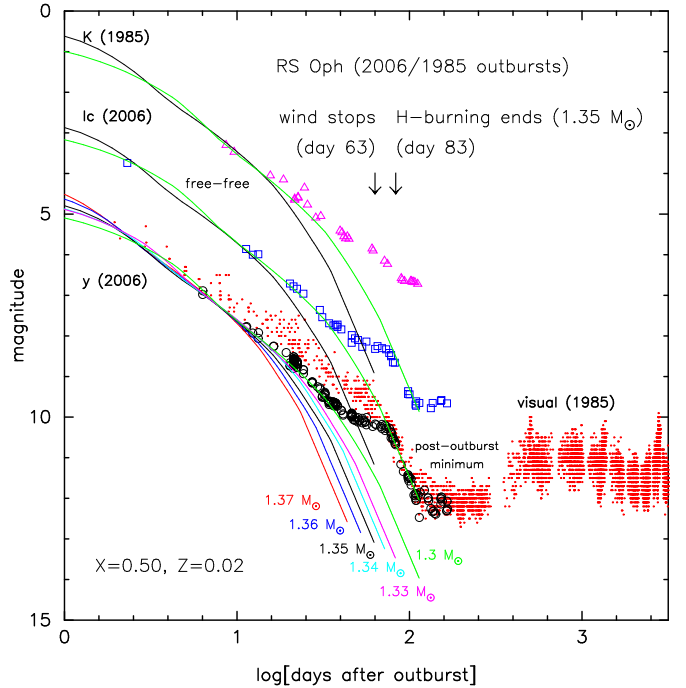}
\caption{
Calculated light curves for free-free emission during the optically
thick wind phase together with observational $y$- ({\it open circles})
and $I_c$-magnitudes ({\it open squares}) of the 2006 outburst.
We also add observational infrared $K$- \citep[{\it open triangles}:
taken from][]{eva88} and visual magnitudes ({\it small dots}: taken from
AAVSO) of the 1985 outburst.
For the $y$-magnitudes, we show six light curves for different WD masses,
i.e., $M_{\rm WD}= 1.3$, 1.33, 1.34, 1.35, 1.36, and $1.37~M_\sun$ 
from right to left.
For the $I_c$- and $K$- magnitudes, we show two light curves 
for two WD masses, i.e.,
$M_{\rm WD}= 1.3$ and $1.35~M_\sun$ 
from right to left.  Here we assume $X=0.5$ and $Z=0.02$ for the 
chemical composition of the envelope.
\label{mass_v_uv_x_rs_oph_x50z02}
}
\end{figure}

\subsection{Photospheric evolution of the white dwarf}
\label{optically_thick_wind}

     After a thermonuclear runaway sets in on a mass-accreting WD,
its photosphere expands greatly and an optically thick wind massloss
begins.  The decay phase of novae can be followed by a sequence
of steady state solutions \citep[e.g.,][]{kat94h}.
After the optically thick winds stop,
the envelope settles into a hydrostatic equilibrium where its mass
is decreasing by nuclear burning.
When the nuclear burning decays, the WD enters a cooling phase, in
which the luminosity is supplied with heat flow from the ash of
hydrogen burning.
We have followed nova evolution,
using the same method and numerical techniques as in \citet{kat94h}.


\begin{figure}
\epsscale{1.20}
\plotone{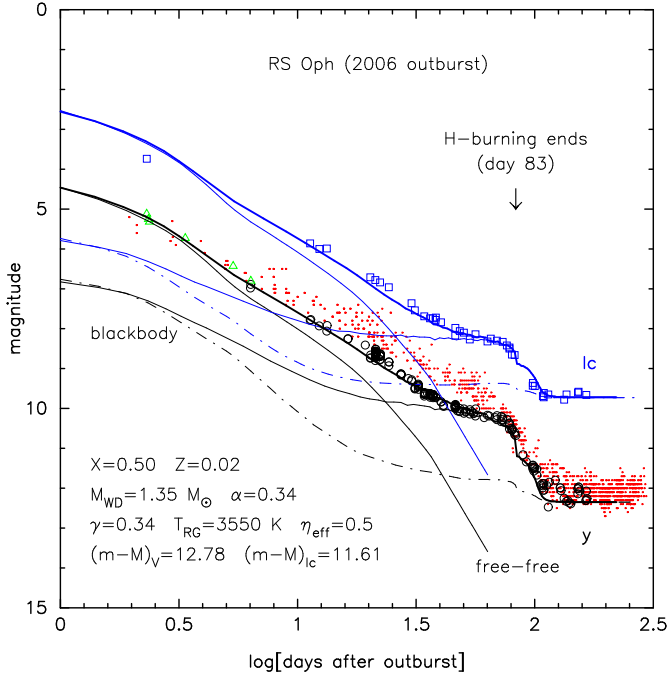}
\caption{
Same as Fig. \ref{mass_v_uv_x_rs_oph_x50z02}, but for a combination of
our free-free plus disk irradiation models.  The plateau phase
is reproduced with a relatively large disk, $\alpha = 0.34$,
for the 50\% iraddiation efficiency.  The various parameters
are summarized in Table \ref{parameter_of_rsoph}.  {\it Thin solid lines}
denote the free-free emission brightness (labeled by ``free-free'')
or the blackbody emission luminosity (labeled by ``blackbody'').
{\it Thick solid lines} are the total of the free-free plus blackbody.
{\it Dash-dotted lines} are the blackbody luminosity only from
the WD and the RG with irradiation.  Several $V$-magnitudes in the very
early phase of the 2006 outburst ({\it Open triangles}) are added.
We cannot reproduce the mid-plateau phase without a disk.
The final sharp decline corresponds to the end of the hydrogen
shell-burning.
\label{nuclear_burning_rs_oph_x50z02_m1350_wide}
}
\end{figure}

In our nova light curve model, we assume that free-free emission of
the optically thin ejecta dominates the continuum flux in the early
phase of RS Oph outbursts as in many classical novae \citep[e.g.,][]{gal76}.
This is the main and most important 
difference from previous Hachisu \& Kato's (2001b) model,
in which the blackbody emission is assumed.

The free-free emission of optically thin ejecta is estimated by
equation (9) in \citet{hac06kb}.
The calculated free-free light curves are shown in Figure
\ref{mass_v_uv_x_rs_oph_x50z02}.  
The decline rate of the light curve, i.e., 
the evolutionary speed depends very sensitively on the WD mass if 
its mass is very close to the Chandrasekhar mass 
\citep[e.g.,][]{kat99, hac06ka, hac06kb}.  This is because
the WD radius is very sensitive to the increase in mass near
the Chandrasekhar mass.

The timescale also depends weakly on the chemical composition of envelopes.
Hydrogen depletion is expected.  This is because, just after
the unstable nuclear burning sets in, convection widely develops
and mixes processed helium with unburned hydrogen.  This mixing reduces
the hydrogen content by 10\%$-$20\% for massive WDs
like in \object{RS Oph}.  The CNO abundance is not
enriched in the recurrent novae, so we adopt the hydrogen content of
$X=0.50$ and the solar metallicity of $Z=0.02$.

We added infrared $K$-magnitudes of the 1985 outburst
observed by \citet{eva88}.  Very little dependence of the light
curve shape on the wavelength is a characteristic feature in the 
free-free emission light curves.  However, the free-free light curve 
cannot reproduce the mid-plateau phase, so we introduce an irradiated
disk in the next subsection.

\subsection{The irradiated disk and companion}
We assume an axi-symmetric disk with a size of
\begin{equation}
R_{\rm disk} = \alpha R_1^*,
\label{accretion-disk-size}
\end{equation}
and a thickness of
\begin{equation}
h = \beta R_{\rm disk} \left({{\varpi} / {R_{\rm disk}}} \right)^\nu,
\label{flaring-up-disk}
\end{equation}
where $R_{\rm disk}$ is the outer edge of the disk,
$R_1^*$ is the effective radius of the inner critical Roche lobe
for the WD component, $h$ is the height of the surface from
the equatorial plane, $\varpi$ is the distance from the symmetry axis,
and $\nu$ is the power index which describes flaring-up of
the disk edge \citep[see, e.g.,][]{sch97, hac01kb}.  We also assume
a companion red giant star with a radius of
\begin{equation}
R_{\rm RG} = \gamma R_2^*,
\label{rad-giant-size}
\end{equation}
where $R_2^*$ is the effective radius of the inner critical Roche lobe
for the red giant component, its mass of $M_{\rm RG}= 0.7 ~M_\sun$,
and the inclination angle of the binary,
$i= 33 \arcdeg$ \citep[e.g.,][]{dob94}.  \citet{dob96} 
suggested $\gamma \sim 0.4$ for the distance of 1.5 kpc.

The disk surface absorbs UV and supersoft X-ray photons from the WD
and emits a part of it as a thermal emission with a lower temperature
than that of the WD photosphere.  The resultant light curve depends
mainly on the disk size ($\alpha$), the efficiency of irradiation 
($\eta_{\rm eff}=$ radiated energy$/$absorbed energy), and also
the radius of the RG, but depends very weakly on the other two
parameters of $\nu$ and $\beta$.  Here, we assume $\nu=2$ and
$\beta = 0.05$.  The dependence on these parameters was widely
discussed in the previous papers \citep[e.g.,][]{hac01kb, hac03ka}.
The irradiation efficiency of the RG hardly affects the light curve
as shown in Figure \ref{nuclear_burning_rs_oph_x50z02_m1350_wide}.

We have obtained three best fit models of the 2006 outburst
in Table \ref{parameter_of_rsoph}.  The calculated light curves
for the 50\% efficiency are plotted in Figure
\ref{nuclear_burning_rs_oph_x50z02_m1350_wide}.  These models reproduce
the mid-plateau phase and the sharp final decline identified as the end
of hydrogen shell-burning.  The turnoff date of day 83
in our $1.35~M_\sun$ WD model is very consisting
with the supersoft X-ray turnoff on day $\sim 90$ observed with
{\it Swift} \citep{osb06}.





\begin{figure}
\epsscale{1.20}
\plotone{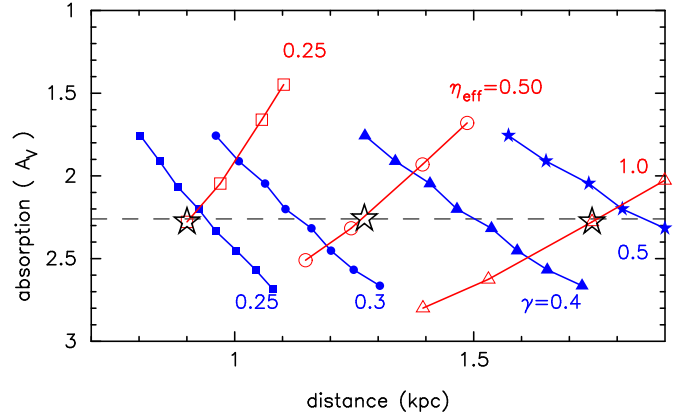}
\caption{
The absorption-distance relation is plotted for the disk irradiation
model and the companion model.  Lines with
{\it open squares}, {\it open circles}, and {\it open triangles}
denote the disk irradiation model with the efficiency of 25\%, 50\%,
and 100\%, respectively.  Each symbol corresponds to the disk size,
i.e., $\alpha= 0.64$, 0.48, 0.32, and 0.24
from top to bottom.  Lines with
{\it filled squares}, {\it filled circles}, 
{\it filled triangles}, and {\it filled stars}
denote the companion star model with the radius of $\gamma = 0.25$,
0.30, 0.40, and 0.50, respectively.
Each symbol corresponds to the photospheric temperature of the 
companion star, i.e., $T_{\rm RG}=3200$, 3300, 3400, 3500, 3600, 3700, 3800, 
and 3900~K from top to bottom.  The absorption of $A_V = 3.1 E(B-V)= 2.26$
is indicated by a {\it dashed line} \citep{sni87}.  Three models in Table
\ref{parameter_of_rsoph} are indicated by {\it large open stars}.
\label{rsoph_absorption_distance}}
\end{figure}





\subsection{Distance}
\label{distance}

     The distance is obtained from the light curve fitting both at
the plateau phase and at the post-outburst minimum phase as shown in
Figures \ref{mass_v_uv_x_rs_oph_x50z02} and
\ref{nuclear_burning_rs_oph_x50z02_m1350_wide}.
The disk luminosity depends mostly on the disk size, $\alpha$,
and the irradiation efficiency, $\eta_{\rm eff}$.  We have changed
these two parameters and calculated the brightness at the mid-plateau
phase.  Fitting the calculated brightness with the observation,
we obtain the apparent distance moduli, i.e., 
$(m-M)_y = (m-M)_V$ and $(m-M)_I$.
Then we calculate the absorption from
\begin{equation}
A_V = {{A_V - A_I} \over {0.518}} = {{(m-M)_V - (m-M)_I} \over {0.518}},
\end{equation}
where we use $A_I = 0.482 A_V$ \citep[e.g.,][]{rie85}.
Once $A_V$ is obtained, the distance is calculated from 
$\log (d) = ((m-M)_V - A_V -5)/5$.  Thus, we obtained
the absorption-distance relation for the irradiated disk
as shown in Figure \ref{rsoph_absorption_distance}.
We further restrict the distance with the observed absorption
of $A_V \sim 2.3$ \citep{sni87}.
The same method is applied to the companion star, in which
we have changed the companion size, $\gamma$, and the 
effective temperature, $T_{\rm RG}$.  The absorption-distance
relation for the companion is also plotted.

   The largest ambiguity of our model is the irradiation 
efficiency of the disk.  Here, the distances of 0.9, 1.3,
and 1.7~kpc are derived for the three different assumed efficiency,
i.e., 25\%, 50\%, and 100\%, respectively.
The actual efficiency is somewhere between 50\% and 100\%, so we 
have a reasonable distance of $1.3-1.7$~kpc. 

\section{Discussion}
\label{discussion}

The decline rate of free-free light curve and the hydrogen burning
turn-off date depend not only on the WD mass but also on the chemical
composition of the envelope \citep{hac06kb}.
We have examined two other cases of
hydrogen content, $X=0.70$ and $X=0.35$, and found that the best
fit models are obtained for $M_{\rm WD}= 1.36$ and $1.34 ~M_\sun$,
respectively.  So we conclude that the WD mass is
$1.35 \pm 0.01 ~M_\sun$.

There are still debates on the distance to RS Oph.
In the previous outburst, \citet{hje86} estimated the distance
to be 1.6 kpc from \ion{H}{1} absorption-line measurements.
\citet{sni87} also obtained the distance of
1.6 kpc assuming the UV peak flux is equal to
the Eddington luminosity.
\citet{har93} calculated a distance of 1290 pc from the
$K$-band luminosity.  \citet{hac01kb} obtained a smaller distance
of 0.6 kpc from the comparison of observed and theoretical UV fluxes
integrated for the wavelength region of 911-3250 \AA~.
This shorter distance is caused by their blackbody assumption, 
because the flux is much larger than the blackbody flux
in this wavelength region.
\citet{obr06} estimated the distance of 1.6 kpc from VLBA mapping
observation with an expansion velocity indicated from emission line width.
\citet{mon06} estimated a shorter distance of $< 540$ pc assuming
that the IR interferometry size corresponds to the binary separation.
If we assume this corresponds to a circumbinary disk, however,
a much larger distance is obtained.  Therefore, our new value
of $1.3-1.7$ kpc is consistent with other estimates.

\acknowledgments

We thank the Variable Star Observing League of Japan (VSOLJ)
and the American Association of Variable Star Observers (AAVSO)
for the visual data of \object{RS Oph}.
This research has been supported in part by Grants-in-Aid for
Scientific Research (16540211, 16540219)
of the Japan Society for the Promotion of Science.

\clearpage

\end{document}